# System and sub-system energy resilience during public safety power shutoffs (PSPS) in California - An evidence-based argument

Dan Thompson, PhD Candidate at University College London (UCL)
Dr. Gianluca Pescaroli, Associate Professor at University College London (UCL)
Dr. Maham Furqan, Remote Fellow at Oxford Institute for Energy Studies (OIES)

*Note: This work is part of a larger project that seeks to understand the relationship between power outage and de-energization data across smaller, discrete parts of the grid which can help identify investment opportunities in behind-the-meter or microgrid solutions in California.*

## Abstract

This study examines historical relationships between Public Safety Power Shutoff (PSPS) events enacted by California's investor-owned utilities (IOUs), at the system and sub-system levels, along with other disruptions to macro electricity systems. This study contributes to understanding the balance between system-wide resilience goals, such as wildfire hazard mitigation, and sub-system-level priorities, such as minimizing the frequency and duration of localized disruptions. Focusing on circuit-level data from 2018 to 2023 as a proxy for sub-systems, we evaluate differences in outage frequency, duration, and customer impact across three major IOUs in California. Results highlight a differentiation between 'higher impact' de-energization events, which have occurred less frequently, and circuits impacted frequently but with lower customer or duration impacts. Study outcomes suggest that resilience, from the perspective of PSPS event, may be more temporal, which in this case is driven infrastructure and planning investments by IOUs. Future work aims to incorporate socio-demographic factors, including urban-rural divide, to identify further opportunities for enhancing resilience at the circuit and sub-circuit levels.

## Acronyms

| Acronym | Term | Acronym | Term |
|---|---|---|---|
| **IOU** | Investor-owned utilities | **SCE** | Southern California Edison |
| **PG&E** | Pacific Gas & Electric | **SDG&E** | San Diego Gas & Electric |
| **PSPS** | Public Safety Power Shutoff | | |

**Keywords:** Public safety power shutoffs (PSPS), electricity resilience, power disruptions, infrastructure planning, de-energization

## 1. Introduction

Research on energy planning has been increasingly focused on 'resilience' to understand and strengthen electricity systems against the cascading effects of power disruptions (Pescaroli and Alexander 2018). Several existing frameworks have advanced arguments that resilience for systems, including electricity systems (i.e., 'electricity resilience' or 'energy resilience'), should be considered as a holistic goal to preserve functioning (Clark-Ginsberg 2016; Dolan 2021). Although not universally defined, several key features of system-level electricity resilience approaches include prioritizing the grid's ability to prevent, respond to, and adapt to disruption events and other threats, which are prioritized over the preservation of power for individual transmission or distribution components (Roege et al. 2014). More recently, system-level electricity resilience has expanded conceptually to include the role of the grid in causing wildfires (Bayani and Manshadi 2023). This can occasionally conflict with resilience goals from the perspective of the sub-grid level or electricity consumer, which often prioritize maintaining a continuous supply of critical electricity, at minimum, during power interruptions (Thompson and Pescaroli 2023).

Studying Public Safety Power Outage(s) (PSPS) in California offers an opportunity to compare system and sub-system resilience goals in the context of expanding resilience goals to include hazard avoidance. This raises two questions: *what does studying a PSPS event from the perspective of the subsystem reveal about the conceptual tension between system and sub-system level resilience goals? Pursuant to this question, how can a sub-system focus on PSPS events be understood from a perspective of other disruptions to the electricity grid, which are much more established?*

PSPS in this case refers to preemptive measures investor-owned utilities (IOUs) take to limit the susceptibility of their electricity equipment to starting wildfires. During a PSPS, the IOUs shut off parts of their distribution and transmission systems when weather conditions—including high winds and low humidity—exceed specified risk thresholds. In 2018, following a series of severe wildfires, the state of California mandated the IOUs to establish processes to enact PSPS, which accompanied mandatory reporting requirements. Utilities were also required to integrate PSPS protocols into wildfire mitigation planning, integrating PSPS into their broader wildfire prevention efforts (Huang et al. 2023; California Public Utilities Commission, n.d.-a). These changes were accompanied by a series of lawsuits following the aftermath of destructive wildfires, which led PG&E to declare bankruptcy (Obeid 2021). Most PSPS have been concentrated in high wildfire threat zones (HFTD Tier 2 and 3), with only 7.6% of all circuits impacted occurring entirely outside these zones from 2018-2023 (analysis from PSPS data; California Public Utilities Commission, n.d.-b). Since 2018, the IOUs have made planning infrastructure changes to mitigate the impact of PSPS (Pacific Gas & Electric, n.d.; Southern California Edison, n.d.; San Diego Gas & Electric, n.d.). This pre-print examines the historical relationship between PSPS events enacted by the IOUs and other disruption events affecting electricity systems.

## 2. Methods

Leveraging a mix of empirical data from California and qualitative data (from stakeholder interviews, policy documents, and available literature), this study investigates how historical PSPS events impact the electricity system in the state and the implications it has for end-users. Based on the size of their service territories, data availability (as the result of mandatory reporting requirements), and their history of investing in smaller resilience solutions over the past 15 years, three major IOUs in California (Pacific Gas and Electric (PG&E), San Diego Gas & Electric (SDG&E), and Southern California Edison (SCE)) were selected.

**Data Analysis**
PSPS data was collected from publicly available sources (California Public Utilities Commission, n.d.-b). After the initial data-cleaning process, we focused on disaggregating data at the circuit level. All data were processed and analyzed in Python. De-energization data was removed if no customers were reported as experiencing de-energization or if the event lasted less than an hour. Any de-energization events that began simultaneously and occurred on the same circuit were treated as part of the same event.

All data were aggregated to the circuit level. The difficulty disaggregating beyond the circuit level to understand which segments of the circuit and, by extension which customers were impacted, during each de-energization event presents a potential limitation in the approach and the interpretation of the results. To mitigate this limitation, we cross-checked the correlation between the total number of sub-circuit disruptions with the total number of times at least one part of the circuit had been de-energized across the reported PSPS timeframes (see time-based clustering) to understand activity at the circuit and sub-circuit level. As a post-processing phase, the IOUs were randomly assigned pseudonyms "Utility A", "Utility B", and "Utility C" in the results and better obfuscate the circuits that experienced higher levels or frequency of de-energization events. Following this cleaning process, the data were segmented into two distinct approaches:

*<u>Frequency-Based Analysis:</u>* This approach was agnostic to the specific timing of circuit de-energization events. Instead, it focused on the frequency of de-energization events for circuits or sub-circuits. The analysis evaluated the frequency against a) the average number of hours circuits or sub-circuits experienced de-energization, and b) the average number of customers affected per de-energization event.

*<u>Time-Based Clustering Analysis (alignment with reported PSPS events):</u>* This approach grouped circuits and sub-circuits based on the time window in which they experienced de-energization events (as defined by the time the utility first de-energized a circuit and the time that it restored the last circuit). Circuit-level data was clustered by day and by utility to align specific de-energization clusters with the outage days reported by the utilities **(Table 1)**.

This method adopted an approach developed for outages caused by disruptions in the bulk power grid across the U.S., specifically examining transmission and generation components (Ekisheva et al. 2021; Ahmad and Dobson 2024; Carreras, Newman, and Dobson 2016). Their approach applied a complementary cumulative distribution function (CCDF; = P(X ≤ x)), which represents the probability that X will be greater than or equal value to a specified value *x*. This approach uses the number of circuits impacted by PSPS events as a function of disruption size and used a log-log function to determine if the CCDF obeyed a power law. To do so, we reframed the scale of analysis of the electricity network from the national-level electricity grid and its relationship with grid-level generation with the IOUs' service territories and their relationship with and the circuit/sub-circuit level (Ekisheva et al. 2021). Elements of the discussion section on the rural/urban characteristics of some of the outlier events, along with mitigation actions that the utilities have enacted from 2018 – 2023, were collected and assessed from publicly available sources (Pacific Gas & Electric, n.d.; Southern California Edison, n.d.; San Diego Gas & Electric, n.d.). The discussion and results section will be expanded using additional documentary data and expert analysis in the full publication.

*Table 1: Reported PSPS event examples (2018)*

| Utility | Date of First De-energization | Date of Last Restoration |
|---|---|---|
| SCE | 12-Oct-18 | 16-Oct-18 |
| PG&E | 13-Oct-18 | 17-Oct-18 |
| SDG&E | 15-Oct-18 | 16-Oct-18 |
| SDG&E | 19-Oct-18 | 20-Oct-18 |
| SCE | 5-Nov-18 | 13-Nov-18 |
| PG&E | 6-Nov-18 | 8-Nov-18 |
| SDG&E | 8-Nov-18 | 9-Nov-18 |
| SDG&E | 11-Nov-18 | 16-Nov-18 |
| SCE | 29-Dec-18 | 1-Jan-19 |

## 3. Results

**Frequency-based analysis:** Results from the frequency-based analysis suggest that the distribution of outages is differentiated between "higher-impact" de-energization events at the circuit and sub-circuit level—in terms of outage hours and customers affected—which occur less frequently, and circuits or sub-circuits that are impacted more often but either for shorter durations or with fewer customers affected. The disaggregation also suggests differences across the IOUs studied. Utility A appears to have significantly fewer repeated circuit or sub-circuit de-energizations, but much higher impacts when these circuits are de-energized. This aligns with some findings on the distribution of outage durations and people affected during grid-wide outages (Kernel distribution; Mukherjee, Nateghi, and Hastak 2018). Most circuits in Utility B seem to have been impacted comparatively less from a customer and hours perspective but were impacted much more frequently over time **(Figure 1)**.

*Figure 1: De-energization events vs hours/customers (2018-2023)*

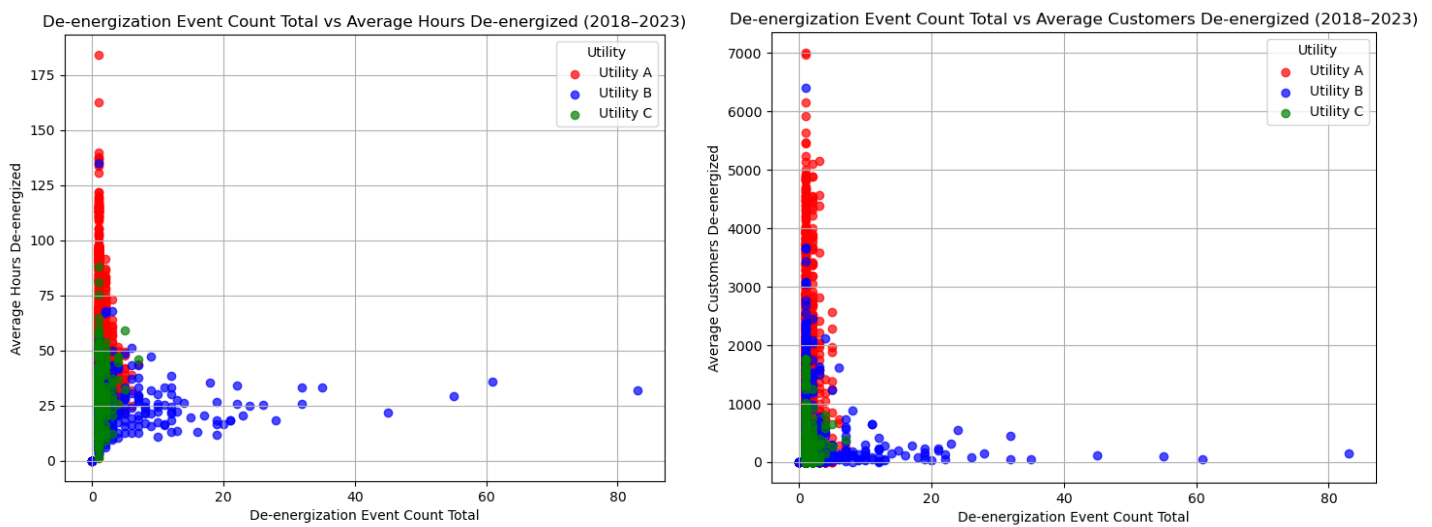

The analysis also combined customers and hours to calculate an average hours-per-customer metric. The results suggested a similar distribution to both metrics shown below; however, this combined metric was not included in the final analysis due to the significantly greater range of average customers de-energized compared to average hours de-energized, which limited the efficacy of using a combination or multiplication of these two variables. Several caveats exist with this data, including the limited timeframe from which it was collected. More importantly, PSPS outages were not evenly distributed across all years studied, as most circuit-level PSPS de-energization events occurred during 2019 and 2020.

*Figure 2: Cluster analysis - circuits impacted vs probability*

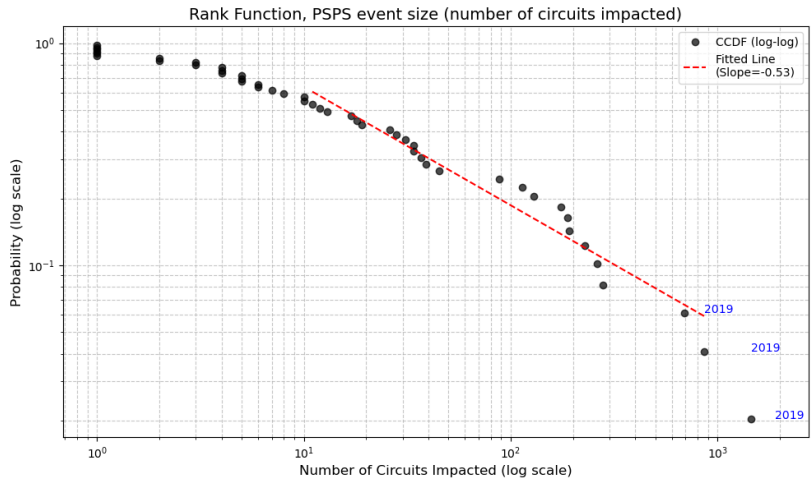

**Time-based cluster analysis:** Results from the time-based cluster analysis **(Figure 2)** suggest that aspects of PSPS events exhibit patterns similar to other disruption events in power systems at the grid level in the middle part of the distribution ($X_{min} = 1$). These results align closely with a power law distribution identified in previous work on grid-level outages. This suggests that the distribution of circuit impacts during PSPS events follows predictable scaling behaviors observed in broader grid resilience studies (Ahmad and Dobson 2024; Carreras, Newman, and Dobson 2016).

The tail decay observed in this study is steeper relative to the previous studies, which adheres to a power law following a minimum cutoff ($X_{min}$) (Carreras, Newman, and Dobson 2016; Ekisheva et al. 2021) and could also represent a lognormal distribution. Testing for fit between the power law and a lognormal distribution, the log-likelihood ratio was negative ($R = -4.1$), suggesting the lognormal distribution provides a slightly better fit; however, the p-value (0.088) is insufficient to reject the power law. Both models are plausible, reflecting the potential for larger-scale de-energization events (in terms of impacted circuits) to be driven by heavy-tailed dynamics. The tail decay may be attributable to a relative lack of data on extreme events. As discussed in the previous section, most of these outages occurred during a peak in 2019 and 2020. More extreme event data will make this relationship clearer. Differences in the results at the tail end of the distribution compared to previous studies also may have resulted from differences in data availability, as some of these studies focused only on the highest-impact events, which may have made the power-law fit more precisely.

Studying PSPS data longitudinally from 2018 – 2023 also suggests that pre-emptive outages resulting from resilience measures (i.e., de-energizations, in this case) could differ from grid outages from hazard impacts. This difference may be attributed to the IOUs' investments in electricity infrastructure and improvements in planning (which we use to refer to monitoring, detection, and communication) over this period, which seem to have limited the impact and potentially the frequency of PSPS events (Pacific Gas & Electric, n.d.; Southern California Edison, n.d.; San Diego Gas & Electric, n.d.; Huang et al. 2023). Resilience, from the perspective of PSPS events, may also be more temporal, reflecting the IOUs' ability to adapt their systems to reduce the impact of PSPS on the number of circuits, customers, and duration of de-energization events over a short timeframe, which contrasts with longer-term, system-wide impacts seen in grid-wide outages. Although utilities have made notable strides in reducing the frequency and duration of PSPS events, our future work in this area, our future work builds on previous studies to suggest that the utilities have not faced aspects of wildfire threat in recent years, particularly large wildfire burns (Huang et al. 2023). We posit that this lack of empirical stress testing on utilities suggests some uncertainty about whether the utilities' current systems are as effective in limiting PSPS impacts.

## 4. Discussion

**Differences in system-level and sub-system-level resilience goals:** Studying PSPS events from a subsystem perspective reveals a conceptual tension between resilience goals at these two levels. System-level goals prioritize the overall grid functionality, including wildfire prevention and hazard mitigation, while sub-system-level goals focus more on localized impacts, particularly minimizing frequency and duration for individual circuits and customers. Our current work geolocates circuits and sub-circuits that have experienced de-energization to understand high-level socio-demographic characteristics, with a particular focus on urban-rural divide and income characteristics. Early results suggest a correlation between the frequency of events and rural areas in HFTD Tier 2 and 3 zones, though these findings are being confirmed through data-triangulation methods.

It is important to note that the IOUs have made notable progress in improving some of these circuits, as demonstrated in their wildfire mitigation reports (Pacific Gas & Electric, n.d.; Southern California Edison, n.d.; San Diego Gas & Electric, n.d.). Nonetheless, the disparity in the data suggests that much of their recent work has focused on limiting major impacts of de-energizations, particularly in terms of average hours lost to de-energization and average customers impacted. In this regard, the IOUs' strategies to adapt components of their system appear logical, as they have concentrated efforts on minimizing higher-impact circuits in terms of customer disruptions and de-energization durations. This may inadvertently overlook a network of circuits that experience higher frequencies of de-energization, raising questions about whether resilience strategies are equally effective across all parts of the IOUs service areas.

**PSPS resilience and comparisons to broader grid disruptions:** The focus on sub-system resilience during PSPS events also offers insights when compared to other electricity system disruptions. Resilience, in response to PSPS events, appears more temporal in nature, reflecting the IOUs' ability to adapt quickly to mitigate impacts. This contrasts with the longer-term disruptions observed during broader grid-wide outages, where recovery and adaptation may take significantly longer. The IOUs' investments in infrastructure and improvements in planning, particularly since 2018, have likely contributed to these temporal gains, allowing them to reduce the frequency and severity of PSPS impacts over time. However, these efforts may also make it challenging to fully evaluate system resilience to conditions that would produce PSPS events, as recent years have not tested the utilities at the same level of wildfire threat levels in 2019 and 2020.

## 5. Conclusion

This study reveals important differences between system-level and sub-system-level resilience goals, particularly in the context of PSPS events in California. While IOUs have made substantial progress in limiting the frequency and severity of higher-impact de-energization events, findings indicate this progress to be uneven. Resilience to PSPS events also appears more temporal compared to other grid-wide disruptions, reflecting the ability of IOUs to adapt more quickly to mitigate the impacts of PSPS, than the impacts of hurricanes on the U.S. grid, for instance (Mukherjee, Nateghi, and Hastak 2018). However, recent conditions have not tested these systems at the same level of wildfire threat observed in 2019 and 2020, leaving uncertainties about their performance under extreme conditions. Our future work will extend this analysis by geolocating circuits and sub-circuits to examine the urban-rural divide, income, and other socio-demographic factors influencing outage frequency and impacts. We will compare these impacts systematically to circuit level infrastructure upgrades reported by the IOUs. These insights aim to help refine strategies for investment in resilience across communities and IOU service areas.